\documentclass[prb,
twocolumn
,showpacs ,floatfix
]{revtex4}
\usepackage{graphicx,amsmath,amssymb,latexsym,amsfonts,subfigure,bbm}
\usepackage[usenames]{color} \usepackage{epic} \usepackage{epsfig}
\newcommand{\x}{\mathbf{x}} 
\newcommand{\vecUV}{\left(\begin{array}[c]{c}u(\x)\\v(\x)\end{array}\right)}

\newcommand{\bge}{\begin{equation}}
\newcommand{\ede}{\end{equation}}
\newcommand{\bga}{\begin{array}}
\newcommand{\eda}{\end{array}}
\newcommand{\Pofl}{\tilde{P}(\tilde{l})}

\begin{document}

\title{Bound states in Andreev billiards with soft walls} 

\author{F. Libisch}
\email{florian@concord.itp.tuwien.ac.at} \author{S. Rotter}
\author{J. Burgd{\"o}rfer} \affiliation{Institute for Theoretical
Physics, Vienna University of Technology, A-1040 Vienna, Austria, EU}
\author{A. Korm\'anyos}
\affiliation{ Department of Physics, Lancaster University, Lancaster,
LA1 4YB, UK, EU}
\author{J. Cserti} \affiliation{Department of
Physics of Complex Systems, E{\"o}tv{\"o}s University, H-1117 Budapest,
Hungary, EU} 
\date{\today}

\begin{abstract}
The energy spectrum and the eigenstates of a rectangular quantum dot
containing soft potential walls in contact with a superconductor are
calculated by solving the Bogoliubov-de Gennes (BdG) equation. We compare
the quantum mechanical solutions with a semiclassical analysis using a
Bohr--Sommerfeld (BS) quantization of periodic orbits. We 
propose a simple extension of the BS approximation which is well suited 
to describe Andreev billiards with parabolic potential walls.
The underlying classical periodic electron-hole orbits are directly 
identified in terms of ``scar'' like features engraved in the quantum
wavefunctions of Andreev states determined here for the first time.
\end{abstract}


\maketitle


\section{Introduction}

The interface between a normal-conducting (N), ballistic quantum dot
and a superconductor (S) gives rise to the coherent scattering of
electrons into holes.\cite{BTK}
This phenomenon, which is of great 
experimental~\cite{Morpurgo,Pothier,Charlat} and
theoretical~\cite{Colin-review, Bagwell, StoneAnd, BenRandMat, MagAndreev,
PseudMagAndreev,Gyorfy,Mortensen,JozNegLen} 
interest, is generally known as Andreev reflection.\cite{Andreev}
A N-S hybrid structure consisting of a superconductor attached to a 
normal cavity is commonly called an Andreev billiard.  
For a recent and comprehensive review of these systems
see, e.g., Ref.~\onlinecite{BenRev} and references therein. 

An important quantity for Andreev billiards is the
density of states (DOS) which has been studied theoretically by many
authors.\cite{Melsen,Lodder,Goldbart,JozQuanMag,JozBoxDisk,
Jozsubgap,JozLogCont,Joz_Sinai,JozCake} Its behavior close to 
the Fermi energy depends on the classical dynamics found in the
isolated normal conducting cavity: In the integrable case, the DOS is 
proportional to the energy, while for a classically chaotic
normal cavity, 
a minigap (which is smaller than the bulk superconductor gap 
$\Delta_0$) develops.
Moreover, depending on the geometry of the normal cavity one can
observe singularities in the DOS. 
A well-known example of such a behavior of the DOS is an Andreev 
billiard formed
from a normal metal film attached to a superconductor 
studied long ago by de~Gennes and Saint-James.\cite{DeGennesArt}
Recently, similar singular features of the DOS have been found in other Andreev
billiards.\cite{JozBoxDisk,Jozsubgap,JozLogCont,Joz_Ferro,JozCake}

To calculate the DOS for Andreev billiards many researchers have 
successfully applied the Bohr--Sommerfeld (BS) approximation.\cite{Melsen,Lodder,Goldbart,Schomerus,Ihra,JozQuanMag,JozBoxDisk,Jozsubgap,JozLogCont,Joz_Ferro,JozCake,BenRev}
It was shown that the DOS can be related to the purely 
geometry-dependent path length distribution $P(l)$ which is 
the classical probability that an electron entering the N region at 
the N-S interface returns to the interface after a path length $l$. 

\begin{figure}[tb]
\epsfig{file=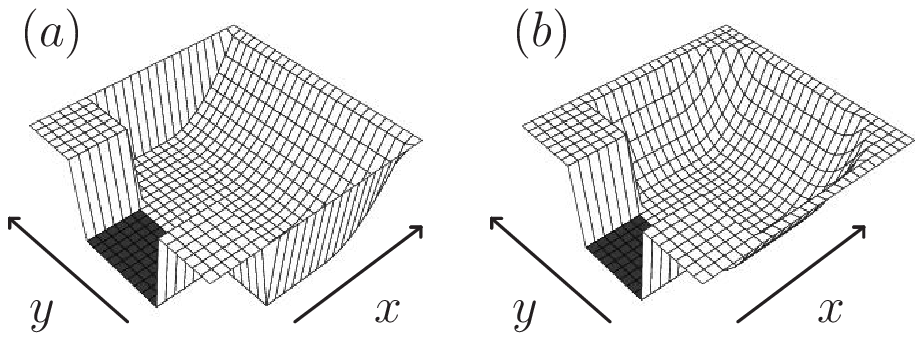,width=8.5cm}
\caption{Andreev billiard with parabolic walls at ($a$) the wall
opposite to the lead and ($b$) at all sides except the one the
superconducting lead is attached to.  
The superconducting area is shaded.}
\label{fig6}
\end{figure}

In the present work, we discuss on the one hand quantum and
semiclassical results for Andreev billiards
with walls which are mediating a smooth transition between the interior of 
the cavity and the region outside.
On the other hand, we calculate, for the first time, the wavefunctions of eigenstates
in such Andreev billiards.
As a prototypical example for an Andreev billiard confined by soft walls,
we choose a parabolic wall profile, as shown in Fig.~\ref{fig6}. 
Such soft-walled Andreev billiards are of interest because they 
describe the potential profile 
found in studies of typical quantum dots realized by remote 
surface gates.\cite{Bird} Soft walls are 
a quite realistic approximation which extends previous theoretical work 
where infinitely high walls were employed.
We demonstrate that with suitable adaptions, the 
BS approximation can describe N-S hybrid systems
 in which the electrostatic potential in the normal conducting cavity 
is modelled by a parabolic potential.
Recently, Silvestrov et~al.~\cite{Silvestrov:cikk} have 
presented a quasiclassical study of Andreev billiards 
containing smoothly varying potentials inside the normal dot.
However, no exact quantum mechanical calculation of the 
energy levels of such Andreev billiards has been performed.  
We succesfully tested the predictions for the DOS obtained from our BS 
approximation by comparing them to results found from exact
quantum mechanical calculations using the Bogoliubov-de~Gennes (BdG) equation.  

A better understanding of the BS approximation can be gained
by the analysis of electron and hole wavefunctions of Andreev
states which we determine by the modular recursive Green's function method 
(MRGM).\cite{StephanDipl,Stephan} Based on recent theoretical
studies~\cite{Colin_vortex,Colin_resonance} the wave function patterns
observed in Andreev billiards may be scanned by measuring the
tunneling conductance of such systems. 
We show that the
wave functions feature enhanced density over continuous families of 
classical periodic orbits (``bundles''\cite{Wirtz}).
These bundles  of classical orbits give rise to peaks in the pathlength 
distribution $P(l)$ and, as a consequence, the DOS shows singular 
behavior at certain energies.
Moreover, a close similarity between the electron and the hole
components of the eigenstates is observed 
in line with the semiclassical picture of 
the Andreev retroreflection process.

This paper is organized as follows. In Sec.~\ref{SecMeth} the method
for calculating the quantum mechanical eigenenergies and eigenstates
is presented along with our extended semiclassical approach. 
In Sec.~\ref{SecRes} our numerical results are presented.
Finally, the paper concludes with a short summary in Sec.~\ref{SecSum}.

\section{Methods}\label{SecMeth}

In this section we outline our quantum mechanical method for determining
the  energy levels and the corresponding 
eigenstates of the N-S hybrid systems shown in Figs.~\ref{fig6}($a$) and ($b$). 
Moreover, we present the semiclassical approach 
for the density of states.

The normal region of the N-S hybrid system is a  
ballistic, rectangular  cavity.  
The length of the side of the cavity 
parallel to the N-S interface is $L$. 
For the spatial dependence of the confining potential $V(\x)$, 
we consider in this work two cases:
i) one parabolic-type soft wall is placed opposite to the N-S interface and
all other sides of the rectangle (apart from the N-S interface) are hard walls  
(see Fig.~\ref{fig6}($a$)), i.e.
\begin{subequations}
\label{soft:VWall-12}
\begin{equation}\label{soft:VWall-1}
V(\x) = \left\{
  \begin{array}{ll} \alpha\, E_F (x-x_0)^2 \Theta\left(x-x_0\right)
 & \rm{if}\ \  0 \leq y \leq L \\  \infty &\rm{elsewhere}
  \end{array}\right. 
\end{equation}
and ii) three sides of the normal cavity are confined by a
parabolic-type soft wall (see  Fig.~\ref{fig6}($b$))
and hence $V(\x)$ is given by 
\begin{eqnarray}
V(\x) &=& \alpha E_{\rm{F}}
\left[ \left(x-x_0\right)^2\Theta(x-x_0)+ y^2 \Theta(-y) \right. 
\nonumber \\
& &  \left.  + \left(y-L\right)^2\Theta(y-L) \right], 
\label{soft:VWall-2}
\end{eqnarray}
\end{subequations}
where in both cases $\alpha$ is a parameter controlling the steepness
of the soft wall, $\Theta$ is the Heaviside function, 
while the parameters $x_0$ and $L$ fix the position of
the potential. The superconducting region of width $W$ ($W \leq L$)
is attached to the remaining side of the cavity. An ideal 
interface between the normal and the superconducting region
is assumed, i.e.~the effective masses and the Fermi energies 
are the same 
in the N and S regions, and no tunnel barrier is present at the interface.   

The superconducting pairing potential $\Delta_0$ is constant in the 
S region and zero in the N region. 
This approximation is valid if  the superconducting coherence length $\xi$ is
small compared to the characteristic size of the rectangular 
cavity.\cite{BenRandMat}
Thus, the pairing potential is given by a 
step function model $\Delta (\x) = |\Delta_0|\ \Theta(x_{\rm{NS}}-x)$, 
where the coordinate system $\x = (x,y)$ is chosen such that 
the N-S interface is located at $x_{\rm{NS}}=0$.

\subsection{Quantum mechanical solution}
 
The quantum mechanical description of the N-S hybrid system is given 
by the BdG equation~\cite{DeGennesArt}
\begin{equation}
\label{Meth:BDG}
\left(\begin{array}[c]{cc}H_0&\Delta\\
\Delta^*&-H_0^*\end{array}\right) \vecUV=\varepsilon\vecUV,
\end{equation}
where $H_0 = {\bf p}^2/(2m^*)+V(\x)-E_{\rm{F}}$ is the single-particle
 Hamiltonian with the confining potential $V(\x)$ that 
defines the normal dot and $m^*$ is the effective mass. 
The electron and the hole part of the quasiparticle wave function are 
denoted by $u(\x)$ and $v(\x)$, respectively, whereas $\varepsilon$ is 
the excitation energy of the quasiparticle measured from the Fermi 
energy $E_{\rm{F}}$. 
In this work we study the bound states of Andreev billiards,
i.e.~the eigenenergies $\varepsilon$ are in the range  
$0 \leq \varepsilon \leq \Delta$. 

The energy levels  $\varepsilon_i$ 
of the hybrid system are found by matching
the wave functions at the N-S interface. 
To construct the wave functions satisfying the Bogoliubov-de Gennes 
equation in the two regions we follow the methods of
Beenakker~\cite{Carlo_PRL:cikk,BenRev}
and Cserti et~al..~\cite{Jozsubgap}
The wave function in the normal region can be expressed in terms of the 
scattering matrix $S(\varepsilon)$ of the open system 
in which the superconductor is replaced by a normal lead. 
This scattering matrix is calculated by  
the modular recursive Green's function method developed  
by Rotter et~al.~\cite{StephanDipl,Stephan}
Note that this method to study Andreev billiards is, 
within the model assumption outlined above,
exact. In particular, it does not rely on the
usual Andreev approximation,\cite{BTK} i.e., $\Delta_0 \ll E_{\rm{F}}$, and 
the assumption of quasi-particles whose angle of 
incidence/reflection are approximately
perpendicular to the N-S interface.\cite{Colin-review}

The integrated density of states (in the following called
 state counting function) can be obtained 
from the energy levels  $\varepsilon_i$ as 
\begin{equation}
  N_{\rm{QM}}(\varepsilon) = \sum_i
  \Theta(\varepsilon-\varepsilon_i).\label{NQM}
\end{equation} 
In our numerical calculations discussed below $N_{\rm QM}(\varepsilon)$
can be obtained directly. 
Therefore, it is straightforward to compare
the state counting functions obtained from the exact quantum
mechanical calculations with that from the BS approximation.
The numerical differentiation to determine the DOS can thus be by-passed.

We now sketch the method for calculating the wave functions
in the two regions.
From the matching conditions one can find the expansion coefficients 
$b_{n}^e\,,\ b_{n}^h$ of the wave functions in the S region, 
and $c_{n}^{e,\pm}\,,\ c_{n}^{h,\pm}$ in the N region 
in terms of the right ($+$) and left ($-$) moving plane waves given by 
\begin{equation}
 \chi_n^\pm(\mathbf{x};E) = \sqrt{\frac{2m^*}{\hbar k_{x,n}W}} 
\ e^{\pm ixk_{x,n}}\sin(y k_{y,n})\label{CHI}, 
\end{equation}
 where 
$k_{y,n} = n\pi/W$ and $k^2_{x,n} = 2m^* E/\hbar^2 -k_{y,n}^2$ 
are the transverse 
and longitudinal wavenumbers respectively, and $W$ is the width of the N-S
interface (here the wave functions are flux normalized).
The wave function in the S region is given by 
\begin{subequations}
\begin{eqnarray}
u_S(\x)
&=& \sum_n \left[ 
\gamma b_{n}^e \chi_n^-(\mathbf{x};E^+)
+\gamma^* b_{n}^h \chi_n^+(\mathbf{x};E^-) \right]\\
v_S(\x)
&=& \sum_n \left[ 
b_{n}^e \chi_n^-(\mathbf{x};E^+)
+b_{n}^h \chi_n^+(\mathbf{x};E^-)\right],
\end{eqnarray}\label{AnsatzS} 
\end{subequations}
where $\gamma = \Delta/(\varepsilon+i\sqrt{\Delta^2-\varepsilon^2})$, 
$ E^\pm = E_{\rm{F}} \pm i\sqrt{\Delta^2-\varepsilon^2}$, and the asterisk 
denotes complex conjugation. Note that to ensure the boundary conditions at
$x \to -\infty$, 
only the right (left) moving plane waves 
$\chi_n^+(\mathbf{x};E^-)$ ($\chi_n^-(\mathbf{x};E^+)$) 
enter in the expansion of the wave function in the S region. 

Calculation of the wave function in the normal dot requires two steps.
At first we obtain the wave function at the N-S interface using the expansion 
coefficients $c_{n}^{e,\pm}$ and $c_{n}^{h,\pm}$
calculated from the matching conditions at the N-S interface.
Secondly, the wave function inside the dot can be written as
\begin{subequations} 
\label{u-v:eq}
\begin{eqnarray}
 u_N(\x,\varepsilon)&=&\sum_n c_{n}^{e,+} \psi_n(\x,\varepsilon),\\ 
 v_N(\x,\varepsilon)&=&\sum_n c_{n}^{h,-} \psi_n(\x,-\varepsilon), 
\end{eqnarray}
\end{subequations} 
where the $\psi_n(\x,\varepsilon)$ are the scattering wave functions of a
particle inside the dot which can be obtained by projecting 
the retarded Green function $G(\x,\x',E)$ of the cavity 
onto the incoming wave:\cite{Baranger}
\begin{equation}
 \psi_n(\x,\varepsilon)= i \hbar \sqrt{v_{x,n}} 
\int_0^W \!\!\!\!\!\!\rm{d}y' \ 
G(\x,\x',E_{\rm{F}} + \varepsilon) \eta_n (y^\prime),  \label{MRGM:WF}
\end{equation}
where  $\eta_n (y)=\sqrt{2/W}\, \sin\left(\frac{n\pi y}{W}
\right)$ and $v_{x,n} = \hbar k_{x,n}/m^*$. 
While the
wavefunction in the superconductor is given as a sum of analytically
determined functions\footnote{Note that the coefficients $b_n^{e,h}$ have to be
determined numerically.} in the continuum limit, the wavefunction in
the normal conductor is determined numerically on a tight-binding
grid. In spite of these two very different approaches, we were able to
fulfill the matching conditions with remarkable accuracy. The latter is a 
measure for the degree of convergence of the tight-binding grid 
calculation towards the continuum limit.

\subsection{Semiclassical treatment}

\begin{figure}[htb]
\epsfig{file=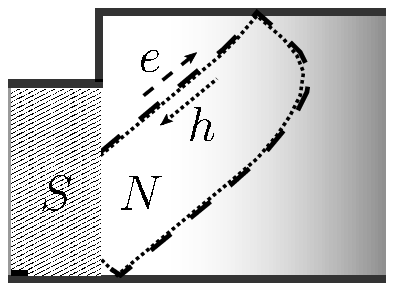,width=4cm}
\caption{A rectangular Andreev billiard with one parabolic wall
  defined by Eq.~(\ref{soft:VWall-1})\ (darker shading in the N region 
corresponds to higher potential). 
One typical Andreev orbit, consisting of an electron part (dashed line) and 
a hole part (dotted line), is shown.
}
\label{figRefl}
\end{figure}

Over the last decade, the Bohr--Sommerfeld approximation for the
smoothed density of states has been successfully applied to Andreev
billiards.\cite{Melsen,Lodder,Goldbart,Schomerus,Ihra,JozQuanMag,JozBoxDisk,Jozsubgap,JozLogCont,Joz_Ferro,JozCake,BenRev}
In the case of a normal dot confined by one N-S interface and infinitely high
 potential walls, the integrated density of states or smoothed state 
counting function in the BS approximation reads
\begin{equation}
\label{NBS}
  N_{\rm {BS}}(\varepsilon) = 
M\sum_{n=0}^\infty\int_{l_n(\varepsilon)}^\infty P(l) dl, 
\end{equation}
where the integer part of $M= k_{\rm F}W/\pi $ 
is the number of propagating modes 
in a lead of width $W$ at the Fermi energy 
$E_{\rm F} = \hbar^2 k_{\rm F}^2/(2m^*)$ 
(here $k_{\rm F}$ is the Fermi wave number)
and $P(l)$ denotes the path length distribution, i.e.~the 
classical probability density for electrons entering the 
billiard at the N-S contact to exit after a path length $l$. 
Finally, $l_n(\varepsilon)$ is given by 
\begin{equation} 
l_n(\varepsilon) = 
\left[n\pi +\rm{arccos}\left(\frac{\varepsilon}{\Delta}\right) \right]
\frac{\hbar v_{\rm{F}}}{\varepsilon}, 
\label{epsqn}
\end{equation}
with $v_{\rm{F}}$ the Fermi velocity. In order to derive Eq.~(\ref{epsqn}), 
one assumes that the hole retraces the path of the electron (retracing approximation, 
see Fig.~\ref{figRefl}). 
This allows one to quantize the periodic orbits created by two subsequent Andreev reflections
using 
the BS quantization condition for N-S systems:
\begin{equation}\label{BScon}
S_e-S_h = \int_{\Gamma}(\mathbf{p}_e-\mathbf{p}_h)d\mathbf{q} =
2\pi \hbar \, (n+\frac{\mu}{4}+\phi).
\end{equation}
In Eq.~(\ref{BScon}) $\mu$ stands for the Maslov index 
(for details see, e.g., Ref.~\onlinecite{Brack}), 
$\phi = \frac{1}{\pi}\, \arccos(\varepsilon/\Delta)$ represents the energy
dependent phase shift resulting from Andreev reflection at the N-S
interface and $\Gamma$ is an arbitrary path of geometric length
$l_g$ connecting one point at the N-S interface with another 
(see e.g.~the one indicated in Fig.~\ref{figRefl}).
The integral in Eq.~(\ref{BScon}) for a normal dot with hard walls results in 
$S_e-S_h =2\varepsilon \, l/ v_{\rm{F}}$  
from which  Eq.~(\ref{epsqn}) follows.

\begin{figure}[htb]
\epsfig{file=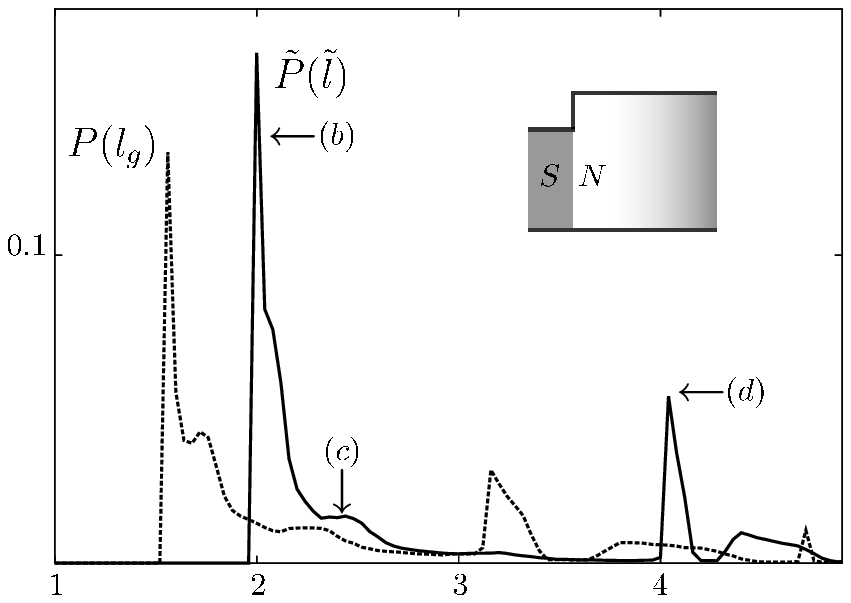,width=7cm}
\caption{Comparison between the geometric path length distribution $P(l_g)$
(taking into account the length of the 
curved trajectories) and the modified path length distribution $\Pofl$
(Eq.~(\ref{sprime})). Both $l_g$ and $\tilde{l}$ are given in units of $L$. The inset shows the 
geometry: Andreev billiard with one parabolic wall, 
$W = 0.7 L$, $\alpha = 6.7$, $x_0 =0.4 L$. The letters 
($b$)-($d$) 
denote the lengths $\tilde{l}$ of those bundles of classical 
orbits, at which the wave functions show enhancement in Fig.~\ref{figVX0.7}.
}
\label{figPall}
\end{figure}

We now extend the BS quantization to soft walls constructed from 
harmonic potentials. Due to the form of the confining potential 
$V(\x) = V(x) + V(y)$ defined in Eq.~(\ref{soft:VWall-12}), the component of 
the momentum parallel to the soft wall is a constant of motion.
Therefore the action (Eq.~(\ref{BScon})) for the part of 
$\Gamma$ which lies in the $V(\x) \neq 0$ region can be decomposed as
$S_e-S_h = S_\parallel + S_\perp$, where $S_\parallel$ 
($S_\perp$) involves the momentum components parallel 
(${\bf p}_{e,\parallel}-{\bf p}_{h,\parallel}$)
and perpendicular (${\bf p}_{e,\perp} - {\bf p}_{h,\perp} $) to the wall. 
Calculation of  $S_\parallel$  is trivial since 
${\bf p}_{e,\parallel}-{\bf p}_{h,\parallel}$ is constant, resulting
in $S_{\parallel}=({\bf p}_e-{\bf p}_h)l_\parallel\sin\vartheta$
where $\vartheta$ denotes the angle between ${\bf p}_{e}$  and 
${\bf p}_{e,\parallel} $ or between  ${\bf p}_{h}$  and ${\bf p}_{h,\parallel}$,
while $l_\parallel $ corresponds to the
displacement parallel to the soft wall which is approached by the particle.
$S_\perp$ can be evaluated for the parabolic potential 
(Eq.~(\ref{soft:VWall-12})) as 
\begin{equation}
  S_\perp =  \pi\, \sqrt{\frac{2m^*}{\alpha E_{\rm{F}}}}\, \varepsilon
  \cos^2  \vartheta .
\label{S_perp:eq}  
\end{equation}
Note that within the retracing approximation the vectors ${\bf p}_e$ and 
${\bf p}_h$ at a given position $\x$ are antiparallel to each other
while their magnitude is different for $\varepsilon \neq 0$, i.e.~$|{\bf p}_e|\neq|{\bf p}_h|$.
Finally, in Eq.~(\ref{BScon}), the integration over that
part of $\Gamma$ that lies in the potential-free region gives
$2\varepsilon \, l/ v_{\rm{F}}$, where $l$ represents the path length of the
electron travelling in the potential-free region.
Putting all these pieces together,  the total integral in Eq.~(\ref{BScon}) 
can be expressed as $S_e-S_h
=2\varepsilon \, \tilde{l}/ v_{\rm{F}} $, where 
\begin{equation}
  \tilde{l} = l + l_\parallel \sin \vartheta + 
  \pi\frac{\cos^2  \vartheta }{\sqrt{\alpha}} \, .
  \label{sprime}
\end{equation}
This result implies that for soft-wall billiards with a confining
potential given by Eq.~(\ref{soft:VWall-12}),
 the Bohr--Sommerfeld approximation
to the state counting function (Eq.~(\ref{NBS})) is still 
applicable if the classical geometric pathlength distribution $P(l)$ is 
modified to account for the potential contribution to the action 
(Eq.~(\ref{S_perp:eq})). The BS quantization therefore involves
a modified pathlength distribution 
$\Pofl$. Accordingly, we perform a Monte Carlo simulation
with typically $10^6$ trajectories, where each 
trajectory is calculated by a forth-order Runge-Kutta
integration. We determine its length $l$ in the potential
free region as well as the contribution from the 
displacement parallel to the soft wall $l_\parallel$. 
Subsequently, $l$ is transformed to $\tilde{l}$ (Eq.~(\ref{sprime}))
for each encounter with one of the soft walls. Note that the 
return probability $\Pofl$ obtained
in this way differs from that calculated from the geometric length of the 
curved trajectories. 
The effective path length $\tilde{l}$ (Fig.~(\ref{figPall}))
associated with the action of the particle in the harmonic potential
is longer than the geometric length $l_g$. The reason for this surprising 
behavior is that while the individual actions 
$S_\perp^e$ and $S_\perp^h$ of the particle and the hole 
propagating in the  parabolic potential
are smaller than $l_g\,p_\perp^0$, with the geometric
length $l_g$ and the constant momentum $p_\perp|_{V=0}=p_\perp^0$,
the difference between the particle and the hole actions 
$S_\perp^e - S_\perp^h$ is larger than $l_g(p_\perp^{0,e}-p_\perp^{0,h})$.

\section{Numerical results}\label{SecRes} 

In this section we compare the state counting function obtained 
from the exact
numerical quantum calculations with that predicted by the 
BS approximation presented above. We consider a ballistic dot confined 
by ($a$) one or ($b$) three parabolically shaped
potential walls as shown in Fig.~\ref{fig6}. 
Moreover, we present both particle as well as hole components
of the wavefunction of eigenstates of these N-S hybrid systems.

\subsection{One soft wall} \label{1-soft}

\begin{figure}[htb]
\epsfig{file=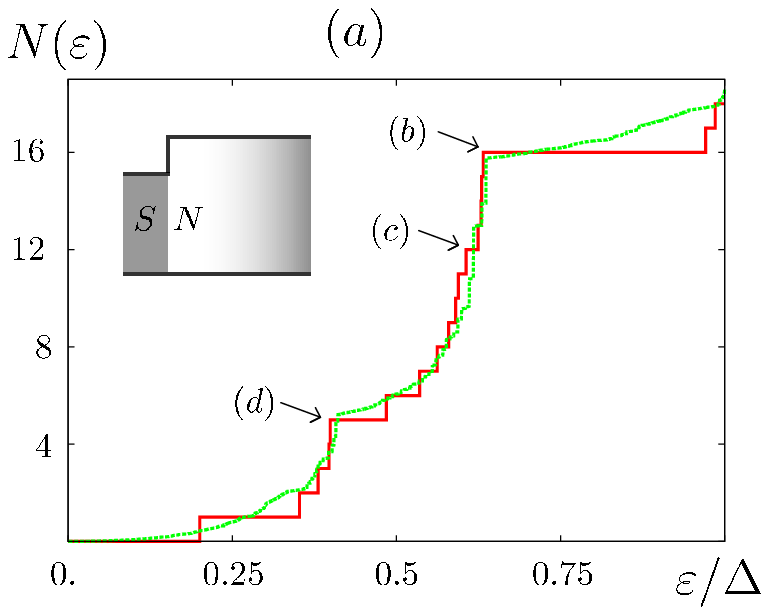,width=8cm}\\
\epsfig{file=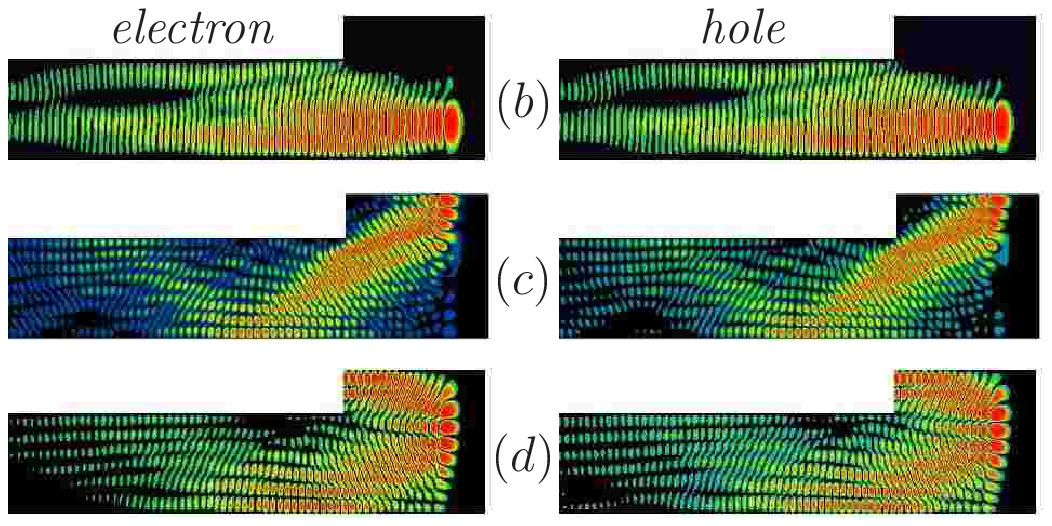,width=8cm}
\caption{(color online) ($a$) Comparison between the quantum
mechanical state counting function $N(\varepsilon)$ (solid red staircase) 
and the BS prediction, $N_{\rm{BS}}(\varepsilon)$, 
(Eq.~(\ref{NBS}), dashed green line) 
for a N-S system shown in the upper left inset.
The modulus square $ |u_N|^2$ and  $|v_N|^2$ (see Eq.~(\ref{u-v:eq})) 
of the three eigenstates marked in ($a$) are shown in ($b$-$d$) 
to illustrate the correspondence between continuous
families of classical trajectories and the quantum mechanical
wavefunctions.
The parameters are $M=15$, $W = 0.7 L$, $h=0$, $\alpha = 6.7$, $x_0 =
0.4 L$ and $\Delta/E_{\rm F}= 0.02 $. 
}
\label{figVX0.7}
\end{figure}

We first consider an N-S system in which the normal dot is confined by
one parabolic potential as shown in  Fig.~\ref{fig6}($a$).  
Numerically we have found that  
most classical trajectories starting from the N-S interface will hit the 
parabolic wall once, before returning to the superconductor (see e.g.~the trajectory 
indicated in Fig.~\ref{figRefl}). This feature
allows us to test our semiclassical approach 
on a comparatively simple level before proceeding to the more complicated 
case of three parabolic walls.
We find that the agreement between the exact quantum mechanical calculations for
$N_{\rm QM}(\varepsilon)$ and its semiclassical counterpart $N_{BS}(\varepsilon)$
is generally very good (see Fig.~\ref{figVX0.7}($a$)), which proves
that  our semiclassical approach extended 
to Andreev billiards with soft walls is well suited to describe these systems.
A small discrepancy  arises only 
at excitation energies above the cusp marked ($b$) in Fig.~\ref{figVX0.7}($a$). 
We also note, that both curves show a distinct cusp structure, 
which has also been found in other 
Andreev billiards.~\cite{JozBoxDisk,JozNegLen,JozCake}
These cusps can be understood from the BS approximation as being the
consequence of the behavior of the path length 
distribution $\Pofl$ at certain path lengths. 
For example, as shown in Fig.~\ref{figPall},
$\Pofl$ has a peak at $\tilde{l}=2 L$ which results in
the most pronounced cusp of the state counting function at 
$\varepsilon/\Delta = 0.63$, as obtained from Eq.~(\ref{epsqn}).
This cusp, marked ($b$) in Fig.~\ref{figVX0.7}($a$), corresponds  
to the shortest classical orbits possible in the system.  
The electron leaves perpendicular to 
the N-S interface, is reflected back at the soft wall, and 
then reaches the N-S interface again. 
Such trajectories are similar to the stationary chords found 
in Ref.~\onlinecite{Goldbart}. The corresponding peak in $\Pofl$ (also marked
as ($b$) in Fig.~\ref{figPall}) represents
a continuous family of trajectories (or bundles\cite{Wirtz}) that 
feature the same topology and bouncing pattern and (almost) the same length,
but different initial conditions. In this case, the bundle is generated by a 
continuous change in the $y$ coordinate at the N-S interface.
The corresponding probability density of the wave function in 
Fig.~\ref{figVX0.7}($b$) 
displays a pronounced enhancement along the bundle 
of classical periodic orbits. 
A similar phenomenon has recently been found in pseudointegrable 
normal billiards~\cite{Bogomolny:cikk} and was called superscarring. 

According to Eq.~(\ref{NBS}), for constant quantum number $n$,
smaller excitation energies
$\varepsilon/\Delta$ correspond to longer classical orbits. 
This suggests that the excitation energies below the pronounced cusp at 
$\varepsilon/\Delta = 0.63$ correspond to classical orbits with 
lengths $\tilde{l} > 2L$.
Indeed, for smaller $\varepsilon/\Delta$ 
certain  wavefunctions mirror bundles 
of longer classical orbits. Some examples are 
 shown in Fig.~\ref{figVX0.7}($c,d$), (see also Fig. ~\ref{figPall} for the correspondig path
lenghts).
The ``superscarring'' of the corresponding wavefunction near the
bundles of short orbits also suggests 
that the BS approximation reflects the 
essential features of the underlying quantum dynamics.

\subsection{Three soft walls} \label{3-soft}

\begin{figure}[!t]
\epsfig{file=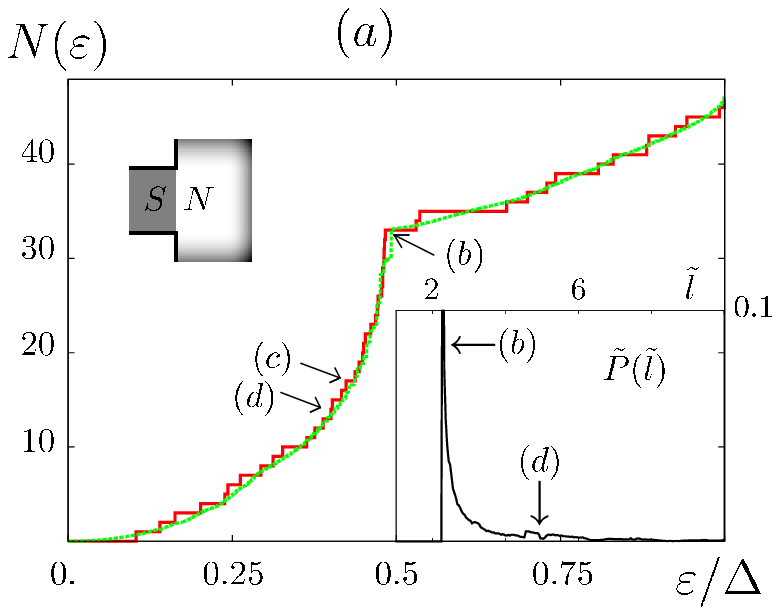,width=8cm}\\
\epsfig{file=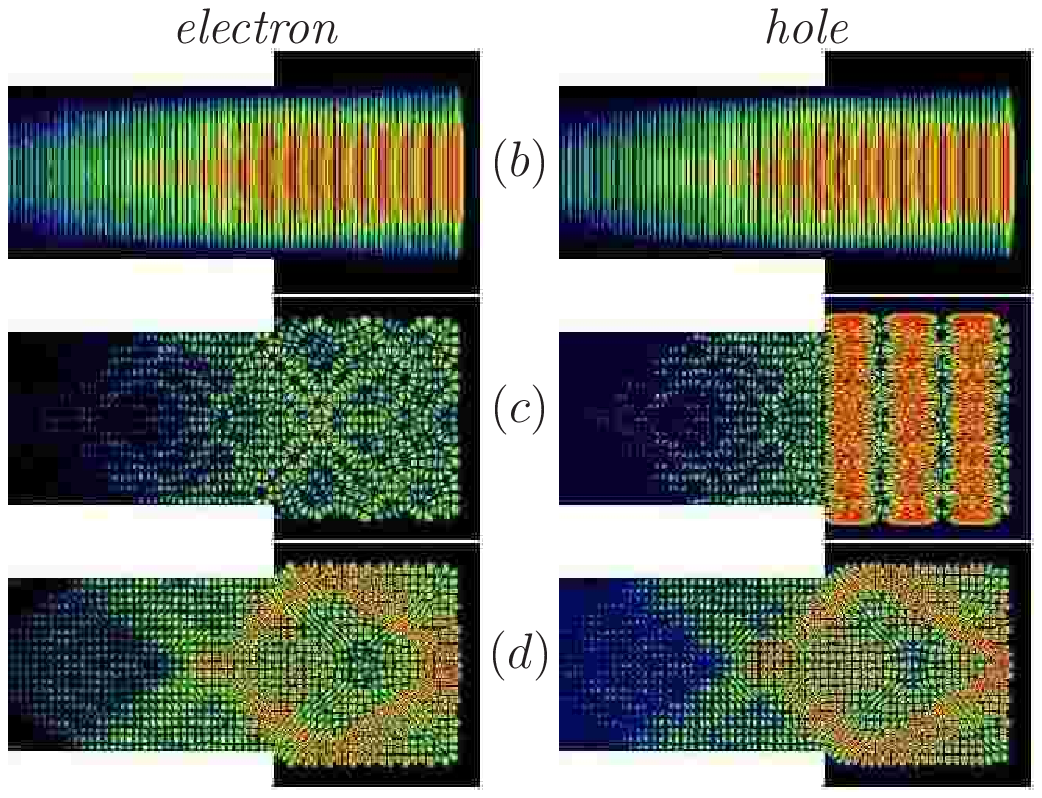,width=8cm}
\caption{(color online)
($a$) Quantum mechanical state counting function (red solid staircase) 
and BS approximation (green dashed line) for 
$\alpha = 5.5$. The parameters are $M=30$, $W = L$,  $\alpha = 125$, $x_0 = L$ 
and $\Delta/E_{\rm F}= 0.02 $. $(b)-(d)$ show selected eigenstates (see text).
}
\label{figVXY}
\end{figure}

We now consider the more complicated case of N-S systems in which 
the normal dot is confined by
three soft walls as shown in Fig.~\ref{fig6}($b$). 

The agreement between the semiclassical BS
approximation (Fig.~\ref{figVXY}($a$)) 
and the quantum mechanical calculation for the state counting function
is again very good. The position of the cusp in
$N_{\rm QM}(\varepsilon)$ can be predicted by the BS approximation 
to originate from the bundle of shortest trajectories. 
The corresponding peak in $\Pofl$ (shown in the lower right inset of 
Fig.~\ref{figVXY}($a$)) is located at $\tilde{l}=2.7 L$ resulting in a cusp at 
$\varepsilon/\Delta = 0.49$. The wave function of the energy level at the cusp 
(see Fig.~\ref{figVXY}($b$)) shows a density enhancement 
over a  bundle of 
classical periodic orbits analogous
to the one discussed
for the dot with only one soft wall. 
As the shortest orbits do not explore 
the upper and lower horizontal walls, this similarity is to be expected. 

Fig.~\ref{figVXY}($d$) presents projections of an Andreev state that explores 
the additional soft walls. It mirrors in both the particle and hole 
components a bundle of classical trajectories all bouncing in sequence at all 
three
soft walls. In line with the retracing approximation, their length 
$\tilde{l}=3$ is reflected in $\Pofl$ (marked as ($d$)). The ``width'' of the 
bundle (i.e.~the continuous range of initial conditions) is, however, narrower 
than the bundle discussed in the previous paragraph,
resulting in a strongly reduced slope 
in $N_{\rm{BS}}(\varepsilon)$.

A very different scenario emerges for the Andreev eigenstate depicted in 
Fig.~\ref{figVXY}($c$). The particle and hole parts display a drastically 
different pattern pointing to the limitation of the ideal
retracing approximation.
While the hole part $v_{\rm N}(\x)$ is characterized by the
``bouncing ball'' type scar of trajectories bouncing at the upper and lower 
horizontal wall, the electron part $u_{\rm N}(\x)$ shows no distinct enhancements. 
Furthermore, integrating  $|u_{\rm{N}}(\x)|^2$ and  $|v_{\rm{N}}(\x)|^2$
over the area of the normal dot we find that the probability of finding the quasiparticle
in the hole state is 79\% while the probability of the electron state is only 12\% 
(and altogether 9\% of finding the quasiparticle in the superconductor).   
Taking into account also that the hole component 
$v_{\rm{N}}(\x)$ has a very low amplitude at  the N-S interface, one can come to the 
conclusion that the hole part is only weakly coupled to the 
superconductor.
The Andreev state of 
Fig.~\ref{figVXY}($c$) can therefore be quantized by assuming that the 
particle is quasi-bound in the hole space where it spends most of its time
with only infrequent and short excursions into the electron space,
where it travels on trajectories scattering off the soft walls before returning
to the hole space. 
 This observation 
suggests that for certain individual
Andreev states, semiclassical EBK quantization becomes possible,
complementing the BS approximation for the smoothed state counting
function.
The  almost vanishing density of
$|v_{\rm{N}}|^2$ at  the N-S interface allows us to assume that the
wave function satisfies hard wall boundary conditions along the
left wall of the cavity.
Accordingly, the eigenenergies are to 
leading order determined by the quantized actions in the hole space in analogy
to long-lived shape resonances. 
The EBK quantization conditions for the ``isolated'' hole state
therefore read
\begin{subequations}
\begin{eqnarray}
2 p_x W_x + \frac{\pi p_x^2}{2\sqrt{2m^* \alpha E_F}}  &=&
  2\pi\hbar \left(n+\frac{\mu_1}{4}\right)\label{Snorm1}\\
2 p_y W_y + \frac{\pi p_y^2}{ \sqrt{2m^* \alpha E_F}}  &=&
   2\pi\hbar \left(m+\frac{\mu_2}{4}\right)\label{Snorm2}
\end{eqnarray}
\label{BS-hardwall:eq}
\end{subequations}
where
$W_x$ and $W_y$ are the linear dimensions of the rectangular region where
$V=0$, while $p_x$ and $p_y$ are the $x$ and $y$ components of the
momentum of the particle. Here $n$ and $m$ are the quantum numbers 
characterizing the actions on the
tori and $\mu_1$ and $\mu_2$ are the Maslov indices.
The left hand sides of these two equations describe the classical actions 
corresponding to the motion of the particle along the $x$ (Eq.~(\ref{Snorm1}))
and $y$ (Eq.~(\ref{Snorm2})) directions.

The Maslov index of $\mu_1 = 3$ in (Eq.~(\ref{Snorm1})) originates from one
reflection at the soft wall and one reflection at the hard wall 
which replaces the N-S interface,
while $\mu_2 = 2$ in Eq.~(\ref{Snorm2}) due to the reflections at
the two soft walls.
The two equations (\ref{BS-hardwall:eq}) can be
solved for $p_x$ and $p_y$ as functions of the 
two quantum numbers $n,m$. The eigenenergy of the N-S system in EBK
quantization is then given by
$\varepsilon = (p_x^2 + p_y^2)/(2m^*)-E_{\rm F}$.
By inspection of Fig.~\ref{figVXY}($c$) we find for 
the hole part of the wave function $|v_N|^2$
three maxima
along the $x$, and 35 along the $y$ direction, respectively. 
Thus, the two quantum numbers in Eq.~(\ref{BS-hardwall:eq}) 
are to be taken as $n=2$ and $m=34$, resulting in 
$|\varepsilon/ \Delta| = 0.47$. 
The excitation energy of the state shown in Fig.~\ref{figVXY}($c$) was 
found to be
$\varepsilon/\Delta = 0.448$. 
This is a remarkable agreement, given that the mean level spacing $\delta_{N}$ 
of the normal dot in units of $\Delta$ is  $0.05$, and 
taking into account  
that semiclassical methods usually cannot, to the leading order, 
resolve the energy spectra on much finer scale than the mean level spacing. 
The small difference between the energies obtained from the BdG equation
and the semiclassical calculation can partly also be attributed to the fact that 
the wave function at the N-S interface is not exactly zero, i.e.,  
a weak coupling of the normal cavity to the superconductor is present. 
 This result implies that the Andreev state in Fig.~\ref{figVXY}($c$)  
is, to a very good approximation, equivalent 
to an eigenstate of the hole in an isolated cavity.
A similar effect of the decoupling of the wave function from the
superconductor has been observed in a system of a superconducting disk
surrounded concentrically by a normal conductor.\cite{JozBoxDisk}
In this hybrid system such states were called whispering gallery states.  

\begin{figure}[hbt] 
\epsfig{file=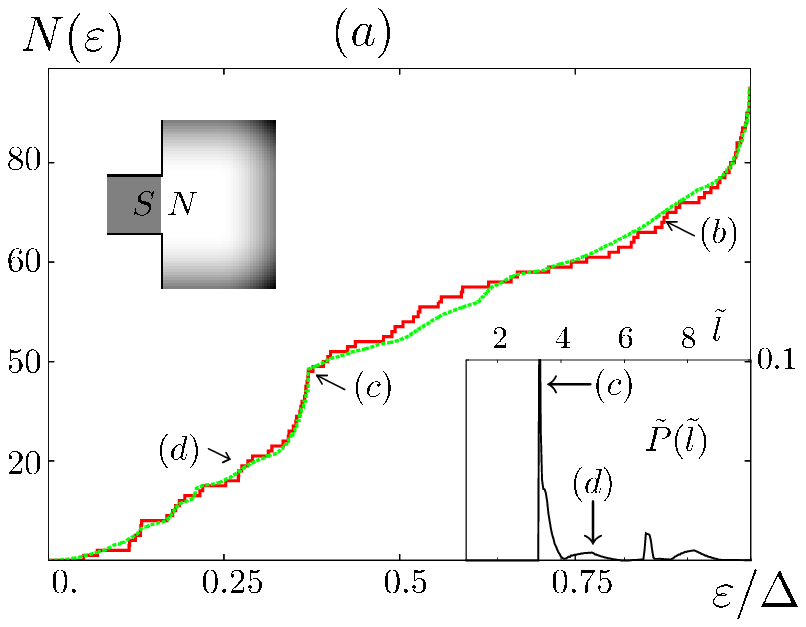,width=8cm}\\
\epsfig{file=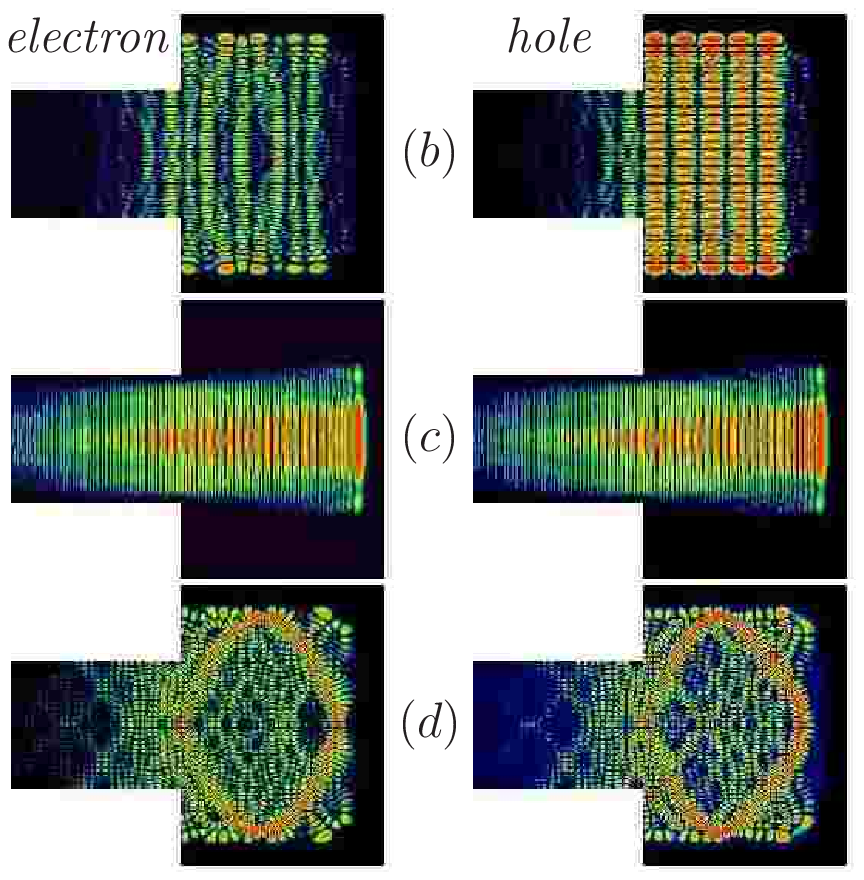,width=8cm}
\caption{(color online) 
($a$) Quantum mechanical state counting function (red solid staircase) and 
BS approximation (green dashed line) for the same system as in 
Fig.~\ref{figVXY} except for a smaller value of 
$\alpha = 5.5$ (curvature of parabolic wall). 
$(b)-(d)$ show selected eigenstates (see text). 
\label{fig:VXYBIG}}
\end{figure}

Finally, we investigate how a smaller $\alpha$, corresponding to
a shallower soft wall, affects the agreement with our semiclassical description.
We consider the same system as in Fig.~\ref{figVXY}, only the parabolic
slope $\alpha$ (or steepness of the wall) is
reduced from $125$ to $5.5$. In such a system, the
particle spends more time in the region where $V \neq 0$. 
The resulting state counting function is shown in Fig.~\ref{fig:VXYBIG}($a$).
The cusp structure is predicted very accurately by our BS approximation.
The most pronounced cusp, marked ($c$)  in Fig.~\ref{fig:VXYBIG}($a$), 
is located at an energy of $\varepsilon/\Delta = 0.37$.
The position of this cusp can again be 
determined by the peak in the path length distribution 
$\Pofl$ at $\tilde{l} = 3.3 L$, corresponding to the bundle 
of shortest trajectories. Note, however, that the non-geometric
shift in length introduced by Eq.~(\ref{sprime}) is $1.3L$, 
and is now comparable to the system size. 

The eigenstate shown in Fig.~\ref{fig:VXYBIG}($b$) is another example of an 
Andreev state which, in a very good approximation, corresponds to a
bound state of the normal conducting billiard. The hole wavefunction
shows 5 maxima in $x$, and 50 maxima in $y$ direction. Inserting
$n=4$, $m=49$ into Eq.~(\ref{BS-hardwall:eq}) yields
an energy of $|\varepsilon/\Delta| = 0.84$. 
The exact energy of the eigenstate shown in 
Fig.~\ref{fig:VXYBIG}($b$) is $\varepsilon/\Delta = 0.86$,
while the mean level spacing of the normal dot for this system is 
$\delta_{N}/ \Delta=0.026$. Note that the mean level spacing is changed as compared
to the previous case where $\alpha=125$,
 because the classically allowed area entering into the expression
of the mean level spacing of a two dimensional dot
is increased.   
 
\section{Conclusions}\label{SecSum}

We investigate an Andreev billiard system with harmonic potential
walls. A quantum mechanical approach using the BdG equation is
presented to calculate the density of Andreev states, and their
wave functions for both the electron and hole component.  
We develop a semiclassical Bohr--Sommerfeld approximation
 for the smoothed state counting function of a soft-walled
Andreev billiard. The quantum mechanical wavefunctions show scar-like
density enhancements which correspond to bundles of semiclassical
Andreev orbits. Additionally, we 
find states which feature very different wavefunctions for electron 
and hole. These states can be understood as quasi-bound states of the
electron or hole component of the normal conducting cavity. Their eigenenergies
can be determined, in a very good approximation, by using
an EBK quantization, assuming
hard wall boundary conditions at the N-S interface.  

\begin{acknowledgments}
 We thank B.~L.~Gy{\"o}rffy and C.~J.~Lambert for helpful discussions.
Support partly by the Austrian Science Foundation (Grant No.~FWF-P17359), the
Hungarian-Austrian Intergovernemental S\&T cooperation program
(Project No.~2/2003), the British Council Vienna,  
the Hungarian Science Foundation OTKA (Grant No.~TO34832) and 
Nanoscale Dynamics and Quantum Coherence (Grant No.~MRTN-CT-2003-504574)
is gratefully acknowledged.
\end{acknowledgments}

\end{document}